\documentclass[journal]{IEEEtran}

\usepackage[pdftex]{graphicx}
\usepackage[caption=false,font=footnotesize]{subfig}
\usepackage{xcolor}
\usepackage{braket}
\usepackage{float}
\usepackage{amsmath}
\usepackage{comment}

\begin{document}

\title{Versatile Filamentary Resistive Switching Model}

\author{Iosif-Angelos Fyrigos,~\IEEEmembership{Member,~IEEE,}
        Vasileios Ntinas,~\IEEEmembership{Member,~IEEE,}
        Georgios Ch. Sirakoulis,~\IEEEmembership{Member,~IEEE,}
        Panagiotis Dimitrakis,~\IEEEmembership{Senior Member,~IEEE,}
        and~Ioannis~G.~Karafyllidis
\thanks{I.-A. Fyrigos, V. Ntinas, G. Ch. Sirakoulis and I.G. Karafyllidis are with the Department of Electrical and Computer Engineering, Democritus University of Thrace, Xanthi, Greece.} 
\thanks{V. Ntinas is also with the Department of Electronics Engineering, Universitat Polit\'ecnica de Catalunya, Barcelona, Spain.}
\thanks{P. Dimitrakis is with Institute of Nanoscience and Nanotechnology, NCSR ``Demokritos", Athens, Greece.}}


\maketitle

\begin{abstract}
Memristors as emergent nano-electronic devices have been successfully fabricated and used in non-conventional and neuromorphic computing systems in the last years. Several behavioral or physical based models  have been developed to explain their operation and to optimize their fabrication parameters. All \textit{existing} memristor models are trade-offs between accuracy, universality and realism, but, to the best of our knowledge, none of them is purely characterized as quantum mechanical, despite the fact that quantum mechanical processes are a major part of the memristor operation. In this paper, we employ quantum mechanical methods to develop a complete and accurate filamentary model for the resistance variation during memristor's operating cycle. More specifically, we apply quantum walks to model and compute the motion of atoms forming the filament, tight-binding Hamiltonians to capture the filament structure and the Non-Equilibrium Green's Function (NEGF) method to compute the conductance of the device. Furthermore, we proceeded with the parallelization of the overall model through Graphical Processing Units (GPUs) to accelerate our computations and enhance the model’s performance adequately. Our simulation results successfully reproduce the resistive switching characteristics of memristors devices, match with existing fabricated devices experimental data, prove the efficacy and robustness of the proposed model in terms of multi-parameterization, and provide a new and useful insight into its operation.
\end{abstract}


\section{Introduction}
\label{Intro}
The main resistance switching (RS) mechanisms in metal-oxide-metal (MOM) memristive devices are very well described in \cite{chua2019handbook,dimitrakis21} and can be categorized \cite{edwards2015reconfigurable} as follows: 1. electrochemical metallization bridge (EMB), 2. metal oxide - bipolar filamentary (MOxBF), 3. metal oxide- unipolar filamentary (MOxUF) and 4. metal oxide - bipolar nonfilamentary (MOxBN). A more compact classification based on the physical mechanisms responsible for the nanoionic nature of the resistive switching \cite{Lanza} includes: the Electrochemical Memories (ECM), the Valence Change Memories (VCM) and the Thermochemical Memories (TCM) \cite{Valov_2011,Valov14}. For the sake of comparison, ECM and EMB are the same, VCM comprises MOxBF and MOxBN while TCM is similar to MOxUF. For all these mechanisms the macroscopic image of the resistance switching is the formation of a conductive filament (CF) that shorts the two electrodes in MOM memory cells, except the case of MOxBN where the transformation of the metal-oxide contact into Schottky occurs. Obviously, MOxBN is the only one where RS depends on the area of the electrodes. The conductive filament may be attributed to metal atoms due to drift/diffusion of metal cations from one electrode to the other (ECM) or to oxygen vacant sites (metal atoms in the oxide bulk with unsaturated bonds) due to drift/diffusion of oxygen vacancies which can be assumed as anions. 

On the other hand, several models have been developed to describe and simulate the operation of resistance switching devices according to the above physical mechanisms. One of the main goals of these models is to describe the phenomenon both quantitatively and qualitatively. Ion drift and diffusion equations are used to simulate the kinetics of ions in the insulator which should be considered in their time and temperature dependent forms. The most common approach to solve the system of these equations is the finite element analysis (FEA) and kinetic Monte Carlo (kMC) \cite{Menzel,Menzel18}. In order to tackle the physics and the computation complexity of the microscopic models, other macroscopic models have been adopted \cite{strukov2008missing,pickett2009switching,waser2009redox,chang2011synaptic,jiang2016compact}. Moreover, various heuristically derived window functions and mathematical techniques have been developed to address the reduced accuracy of the memristor model near the boundaries of device’s kinetics \cite{corinto2012boundary,joglekar2009elusive,biolek2009spice,prodromakis2011versatile,Ntinas}. In parallel, phenomenological models based on the experimental measurements of the RS device behavior have been developed in an efficient way to enable the memristor circuit design \cite{kvatinsky2012team,kvatinsky2015vteam,vourkas2012novel}, compensating accuracy loss and generality with efficient circuit simulations. In brief, the microscopic models are necessary for use in manufacturers process development kit (PDK) and macroscopic models for circuit simulations. 

All of the microscopic models described above do not use quantum mechanical methods to describe and simulate the memristor operation, although both the ionic kinetics and conduction phenomena in the nanoscale can be described accurately using quantum mechanical methods \cite{datta2000nanoscale}. In this work, a filamentary based nanoscale memristor model for ECM has been developed. The model assumes a standard metal-insulator-metal RS device in 2D. The ion kinetics in the dielectric that results in filament formation and deformation are modeled using quantum walks, which incorporate the applied potential distribution in the filament. 

The structure of the filament in the dielectric is described using tight-binding Hamiltonians. The electron transport in the nanoscale filament/dielectric system is modeled using Non-equilibrium Green's Functions (NEGF) \cite{datta2000nanoscale}. Tight-binding Hamiltonians combined with NEGF are used to compute the conductance of the nanoscale memristor as a function of the applied voltage. The process of the filament formation and dissolution in the solid dielectric is modeled by dividing the process into computational time steps. At each step the new filament structure is simulated using quantum walks and the Hamiltonian describing the system is constructed. This Hamiltonian enters the NEGF method to compute the device conductance and current for this step. Our results reproduce successfully the I-V characteristics of RS devices and show that the conductance quantization and the corresponding multiple RS states become more prominent as the solid dielectric thickness decreases. 

The model is fully parametrized and ready for calibration using experimental data. As a proof of concept, experimental results of a fabricated metal-oxide-metal memristive device were used to prove the quantitative and qualitative agreement of the presented simulation results with the fabricated device's ones. Specifically, we fabricated and measured planar MIM devices made of 30nm Cu (active electrode) on a 40nm thick SiO$_2$ deposited by sputtering and 100nm thick Pt (counter electrode). The whole MIM structure was sited on a Ti(5nm)/ thermal-SiO$_2$ (300nm)/n-Si (P-doped, 1-5 $\Omega$cm). Current - voltage sweeps were performed in order to investigate the \texttt{SET} and \texttt{RESET} switching processes. The experimental results suggested a clear bipolar resistance switching operation with \texttt{SET} and \texttt{RESET} voltages~4.5V and -4.5V, respectively. The Cu/SiO$_2$ is a well-known combination to achieve ECM (MOxBF) ReRAM devices \cite{Tappertzhofen}. The resistance switching is attributed to the oxidation of Cu atoms lying at the interface with the oxide and the resulted Cu ions are drift and diffuse across the bulk oxide layer towards the counter electrode. This process has been experimentally proven and modeled by many investigators \cite{Valov14,Menzel15,Valov_2017,Valov_2013,Mehonic}. Furthermore, our model can reproduce successfully the resistive switching characteristics of MOxBF 
RS devices under the appropriate modification of specific parameters. Such a quantum mechanical model can be used to simulate the operation of nanoscale memristors and provides a new and useful insight into their operation.

\section{Model Structure and Outline}
\label{Model}
\begin{figure}[t]
\centering
  \includegraphics[width=0.8\linewidth]{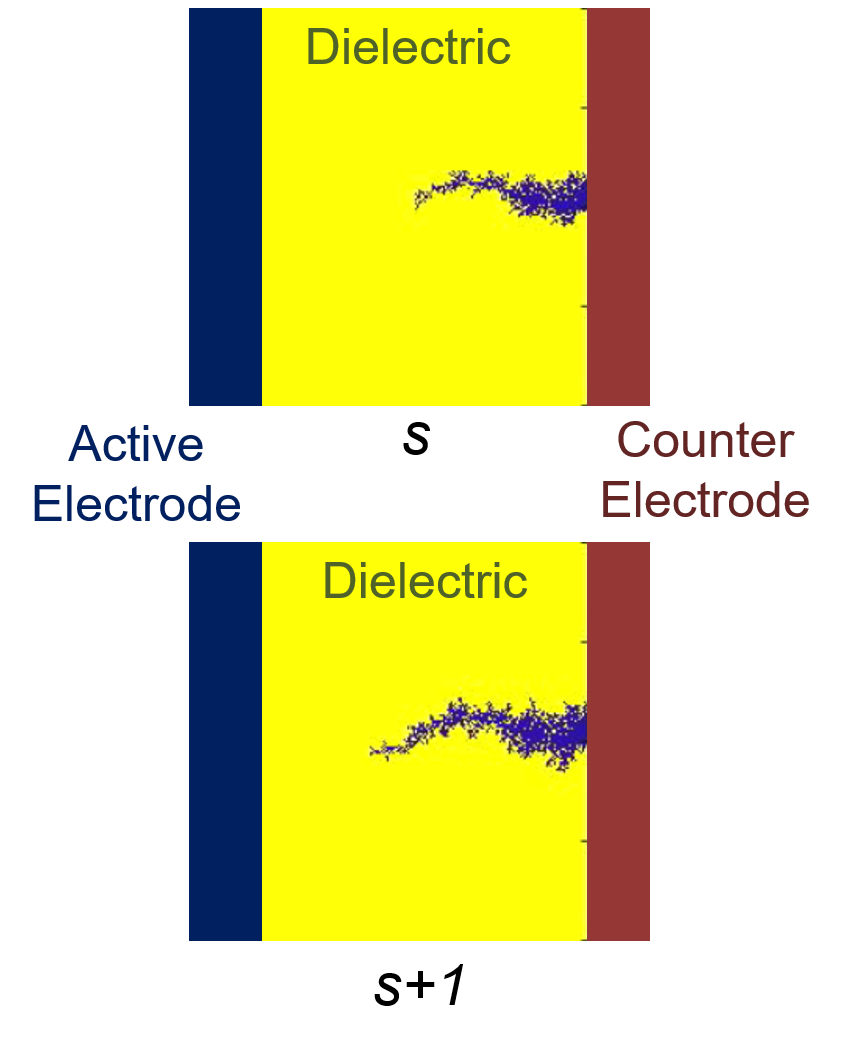}
    \caption{Filament formation in the dielectric at two successive computational time steps, $s$ and $s+1$.}
    \label{Fig:1}
\end{figure}

During the memristor operation a filament is progressively formed in the dielectric, between the active electrode and the counter electrodes. The process of filament formation is simulated in discrete computational time steps. At each step the applied voltage at the device electrodes is increased by a constant value that is the voltage step of the applied voltage sweep. In this way, the dependence of the filament formation process on the voltage sweep rate can be investigated. Figure \ref{Fig:1} shows the progress of filament formation in the memristor at two successive computational time steps.

The filament structure at computational time step $s$ is shown in the upper part of Fig. \ref{Fig:1}. The simulation for computational time step $s+1$ involves the following computation steps, which are briefly provided here and will be thoroughly explained later through the corresponding sections:

\begin{enumerate}
  \item The structure of the filament and the potential distribution in the filament at step $s$, are the inputs for the computations for computational step $s+1$.
\item The filament structure and potential distribution in the filament at computational time step $s$, enters the quantum walk algorithm, in a manner to be explained later on. The quantum walk algorithm computes the kinetics of the metal ions and outputs the filament structure at computational time step $s+1$, shown in the lower part of Fig. \ref{Fig:1}.
\item The applied voltage is increased by a constant value and the potential distribution in the filament structure at computational time step $s+1$ is computed.
\item The tight-binding Hamiltonian \textbf{for the dielectric-filament system} at computational time step $s+1$ is computed. The computation of the Hamiltonian will be explained later on.
\item The potential distribution at computational time step $s+1$ and the Hamiltonian enters the NEGF method, in a manner explained later on, and the conductance of the memristor device at time step $s+1$ is computed.
\item The potential distribution and structure of the filament computed for computational time step $s+1$, are used as inputs for the next computational time step, $s+2$.
\end{enumerate}

As the applied voltage increases, the filament progressively is formed until it reaches the active electrode. After that, the applied voltage is decreased, the metallic ions dissolve in the dielectric and move towards the active electrode, so the filament deforms. Again the quantum walk is used to simulate the kinetics of the filament deformation. The model described briefly above computes the memristor conductance as a function of the applied voltage. In the next sections, as already mentioned, the application of the quantum walk model and the NEGF method in the specific problem will be described analytically. 

\section{Conductance Calculation of Device Utilizing NEGF Method}
\label{Conductance}
The memristor dielectric is divided into a matrix of atomic-size square cells. During the memristor operation these cells are occupied either by atoms of the dielectric or metallic ions, diffusing from the active electrode. Electrons are transported from one electrode to the metallic ions in the dielectric and from there to the other electrode. In the tight-binding approach electrons are transported stepwise from a cell to its neighboring cells at each computational time step. This process is described by the tight-binding Hamiltonian, $H_D$:

\begin{equation}
\begin{array}{l}
{H_D} = \sum\limits_{i,j} {\sum\limits_{k,l =  - 1}^{k,l =  + 1} {( - {t_{i + k,j + l}})} } \hat c_{i + k,j + l}^\dag {{\hat c}_{i + k,j + l}}\; + \\
\;\; + \;\frac{1}{2}\;\sum\limits_{i,j} {\sum\limits_{k,l =  - 1}^{k,l =  + 1} {\hat c_{i + k,j + l}^\dag {{\hat c}^\dag }_{i + k,j + l}\;} } V\;\hat c_{i + k,j + l}^{}{{\hat c}^{}}_{i + k,j + l}
\end{array}
\label{eq1}
\end{equation}

In the general case, an electron is located at the ($i,j$) cell. The electron motion is described using fermion annihilation and creation operators, namely $c$ and $c^{\dagger}$, which annihilate and create electrons in the $(i,j)$ cell and its neighbors $(i+1,j), (i-1,j), (i,j+1)$ and $(i,j-1)$. Electron motion is associated with a kinetic energy $t_{i,j}$, which is the overlap integral. The overlap integral between non-neighboring cells is zero. The overlap integral also describes the effect of temperature on electron mobility. The second term of \eqref{eq1} is the retention energy associated with the tendency of electrons to remain in the current cell, the potential in which is described by $V$.

The conductance of the memristor for various electron energies, $G(E)$, is given by the matrix equation \cite{Moysidis}:

\begin{equation}
G(E)=\frac{2q^2}{\hbar}Trace[\Gamma_1G_D\Gamma_2G_A],
\label{eq2}
\end{equation}

\noindent $G_D$ is the retarded Green's function, and $G_A$ is the advanced Green's function. $\Gamma_1$ and $\Gamma_2$ are the broadening functions of the two contacts: 

\begin {equation}            
\Gamma_1=i[\Sigma_1-\Sigma^{+}_1], \quad \Gamma_2=i[\Sigma_2-\Sigma^{+}_2].
\label{eq3}
\end{equation}

\noindent The contacts are described by their self-energies, $\Sigma_1$ and $\Sigma_2$, respectively. The retarded Green's function is given by \cite{Moysidis18}:

\begin{equation}
G_D=[E-H_D-\Sigma_1-\Sigma_2]^{-1},
\label{eq4}
\end{equation}

\noindent while the advanced Green's function is:

\begin{equation}
G_A=(G_D)^{+}.
\label{eq5}
\end{equation}

\begin{figure}[!t]
\centering
  \includegraphics[width=1\linewidth]{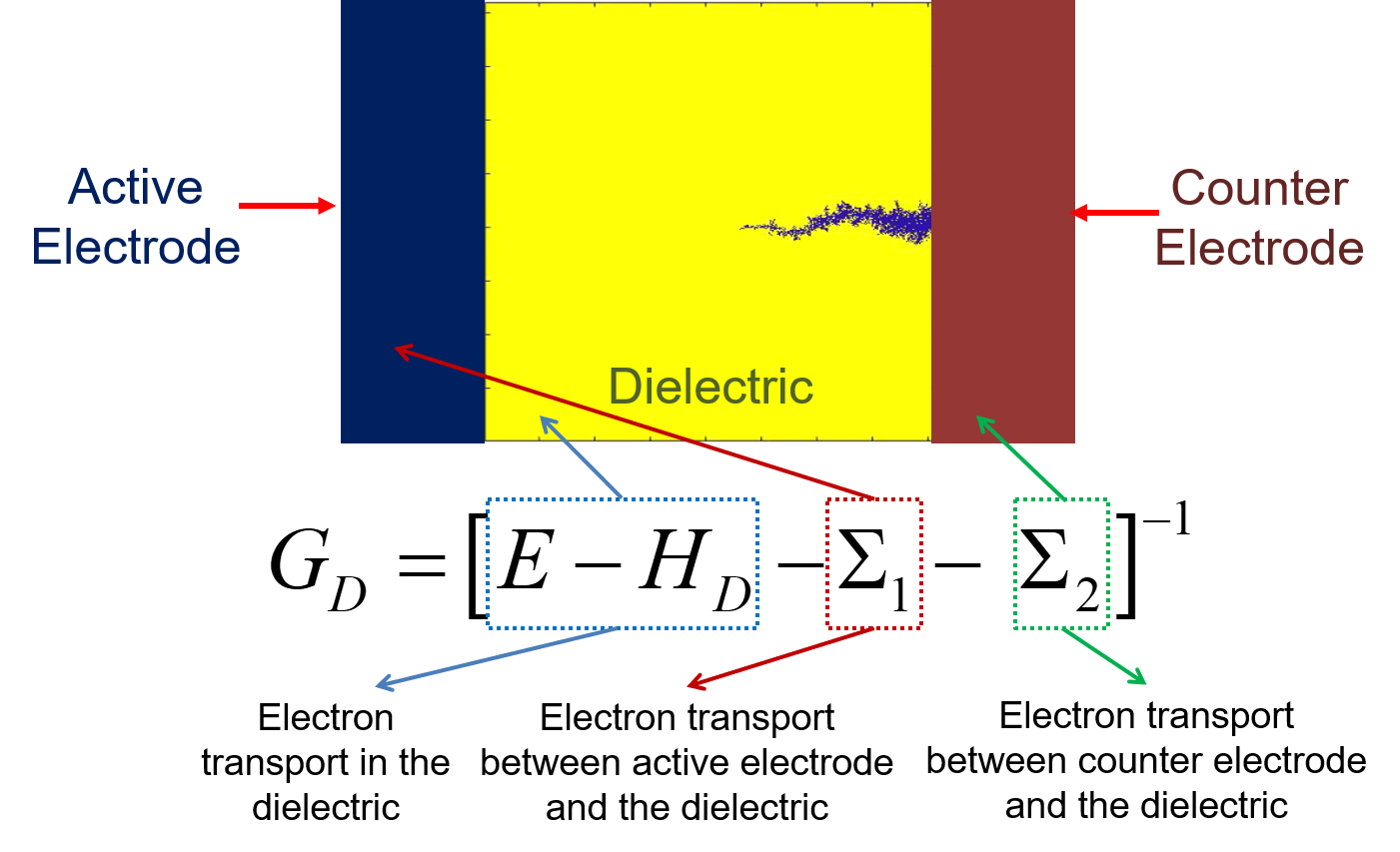}
    \caption{The parts of the matrix eq.~\ref{eq4} for retarded Green's function $G_D$ that describe the active electrode, the dielectric containing metallic ions and the counter electrode, respectively.}
    \label{Fig:2}
\end{figure}

Fig.~\ref{Fig:2} depicts the outline of NEGF method applied to the memristor device. The device is described by the retarded Green's function, which determines its conductance through \eqref{eq5} and \eqref{eq2}. The tight-binding Hamiltonian $H_D$, describes the electron transport in the dielectric-metallic ion system, the self-energy $\Sigma_1$ the electron transport between the active electrode and the dielectric, and the self-energy $\Sigma_2$ the electron transport between the counter electrode and the dielectric. 

\subsection{GPU Acceleration of NEGF}
\label{GPUNEGf}
One of the major issues resulting from the utilization of the NEGF method in the application to the memristive conductance calculations is the request for high performance computing resources. In order to tackle this issue, Graphics processing units (GPUs) originally designed for computer video cards but recently emerged as the most powerful chip in a high-performance workstation, are exploited to boost the performance of the NEGF method. Unlike multicore commercial CPU architectures, 
GPU architectures contains 
thousand of cores capable of running numerous threads in parallel.  As a result, GPU by offering an almost ceaselessly increased number of cores couple with a high memory bandwidth, provides incredible computational resources that enable it 
as a suitable candidate to accelerate NEGF method's computation. 

\begin{figure}[!t]
\centering
  \includegraphics[width=0.99\linewidth]{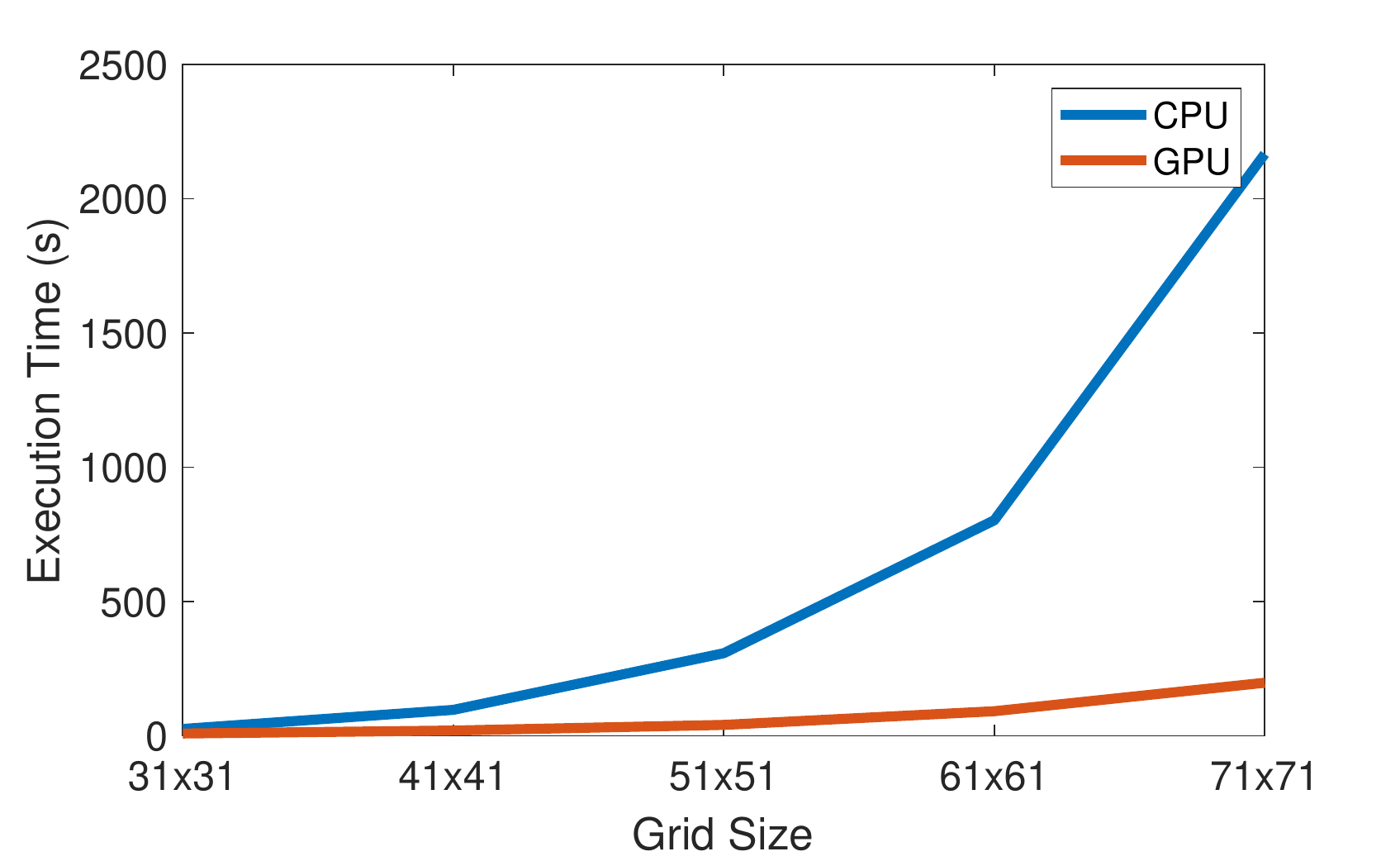}
    \caption{Comparison between CPU (i9-9900K) and GPU (2060 Super) during execution of NEGF method for different number of atomic grids}
    \label{GPU}
\end{figure}

More specifically, NEGF method, when employed to compute the conductance of nanodevices, is a computational demanding technique and conductance becomes harder to be computed as the number of atoms that constitute the device increases. NEGF method parallelization suitability arises from the fact that it contains operations between large matrices which grow exponentially in size as the number of simulated atoms increases. Moreover, the different energy levels ($E$) for which the conductance ($G(E)$) has to be computed are completely independent from each other, meaning that the computation of each energy level can be allocated to a different multiprocessor of the GPU. Therefore, NEGF method was accelerated through GPU parallelization. More specifically, two different algorithms were developed, one for GPUs that are optimized for single precision operations and another, more precise, for GPUs that are optimized for double precision operations, each one suitable for different hardware implementations of GPUs. The results of the comparison as depicted in Fig.~\ref{GPU}, between a up-to-date CPU and a corresponding GPU devices, are very promising showcasing that the NEGF method scales almost linearly in GPU while in the CPU there is an exponential increase of execution time. As a result, the attainable speedup factor of the GPU compared to the CPU was more than $\times{10}$. 

\section{Modelling of Filamentary Growth Sub-system utilizing Quantum Walker's probability distribution}
\label{QuantumWalker}

Due to the nanoscale dimensions of the memristors, the fabricated devices are governed by several quantum phenomena (quantum tunneling, Poole – Frenkel, etc.). To integrate the quantum physics into the modelling of the filament, quantum walks \cite{aharonov1993quantum}, the quantum mechanical counterpart of classical random walks where utilized. The probability of each atomic slot to be occupied by a $Cu$ or $Di$ atom, arise from the motion of the quantum walker.
More specifically, in general, any quantum walker is characterized by its position in the 2D space at any given moment as well as by its quantum state. In our realization the quantum state (coin state) of the walker is described using 2 qubits.
\begin{equation}
|c\rangle= a|00\rangle+ b|01\rangle+ c|10\rangle+ d|11\rangle,
\label{eq10}
\end{equation}

\begin{equation}
|a|^{2}+|b|^{2}+|c|^{2}+|d|^{2}=1,
\label{eq11}
\end{equation}

\noindent where $a$, $b$, $c$ and $d$ are the probability amplitudes, the absolute square of which gives the probability of the walker to be in the corresponding state.

\begin{figure}[!t]
    \centering
      \includegraphics[width=0.7\linewidth]{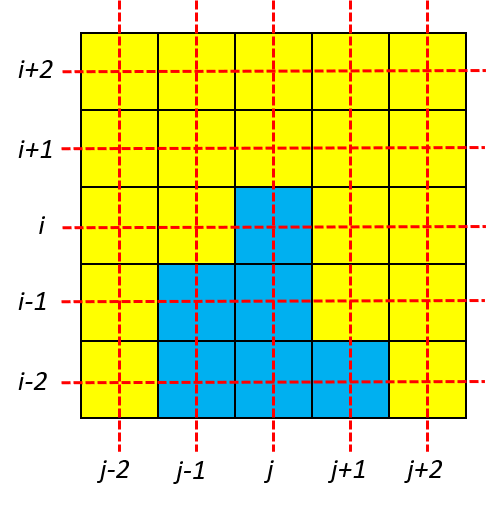}
    \caption{The grid of square cells and the quantum walk lattice. Yellow cells are occupied by dielectric atoms while blue cells by metallic ions. The quantum walk lattice is formed by connecting the centers of the square cells.}
    \label{grid}
\end{figure}

Coin state $|c\rangle$ determines the state of the quantum walker in a specific location. Every position of the 2D grid has a state vector. In the presented case, Fig.~\ref{grid} shows the 2D square grid into which the dielectric-filament system is divided.  Yellow cells represent dielectric atoms and blue cells represent metallic ion atoms. In discrete 2D quantum walk, the walker (metallic ion atom) moves on the lattice formed by connecting the square cell centers, as shown in Fig.~\ref{grid} by red dashed lines. 

The state of the quantum walker at computational time step $s$, namely $|S_s\rangle$, is its position on the lattice at this step:

\begin{equation}
|S_t\rangle=|i,j\rangle.
\label{eq12}
\end{equation}

The Hilbert space of the discrete quantum walk comprises two subspaces, the location subspace $H_L$, which is spanned by the basis:

\begin{equation}
|S\rangle=|-n,n\rangle, L |i-1,j\rangle, |i,j\rangle, |i+1,j\rangle, |i,j+1\rangle, L |n,n\rangle,
\label{eq13}
\end{equation}

\noindent and a four-dimensional coin subspace, $H_C$, which is spanned by the four coin basis states $|00\rangle$, $|01\rangle$, $|10\rangle$ and $|11\rangle$. The Hilbert space, namely $H$, of the quantum walk is the tensor product $\otimes$ of these two spaces:

\begin{equation}
H=H_L\otimes{H_C}.
\label{eq14}
\end{equation}

The state of the quantum walker at computational time step $s$ is the tensor product of its location and its coin state:

\begin{equation}
|S_t\rangle=|\text{location at time } s\rangle\otimes |\text{coin at time } s\rangle.
\label{eq15}
\end{equation}

\noindent For example, the quantum walker's state at location $(i,j)$ with coin in state $|01\rangle$ is $|i,j\rangle \otimes |0,1\rangle$. During the evolution of the quantum walk, the new state of the quantum walker at the next computational time step, $s+1$ is given by:

\begin{equation}
|S_{s+1}\rangle=U(s)|S_s\rangle,
\label{eq16}
\end{equation}

\noindent where $U(s)$ is a unitary operator that drives the quantum walk and it is the product of the following operators:

\begin{equation}
U(s)=P(s)C(s)exp\Big(\frac{i}{\hbar}V(i,j,t)\Big).
\label{eq17}
\end{equation}

\begin{figure*}[!htbp]
    \centering
  \subfloat[]{\includegraphics[width=0.495\linewidth]{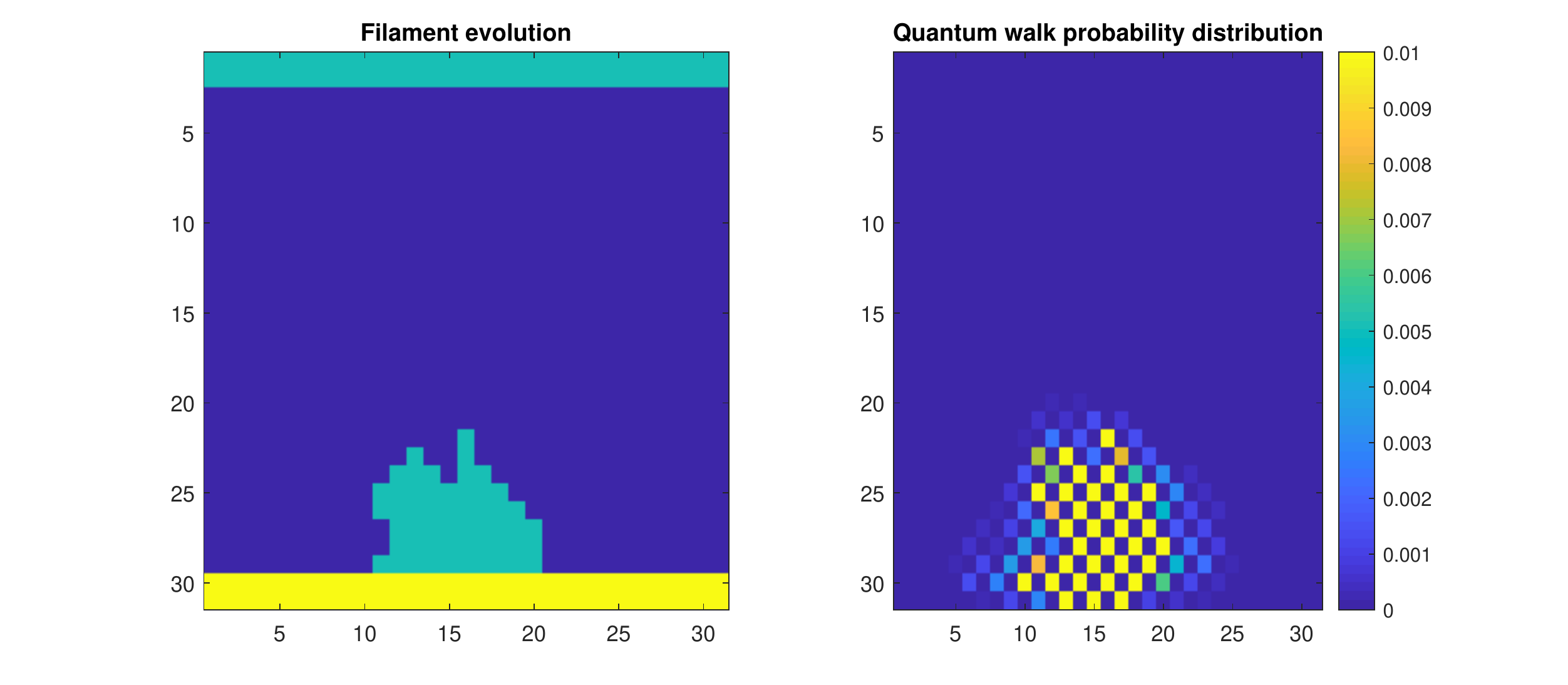}}
  \subfloat[]{\includegraphics[width=0.495\linewidth]{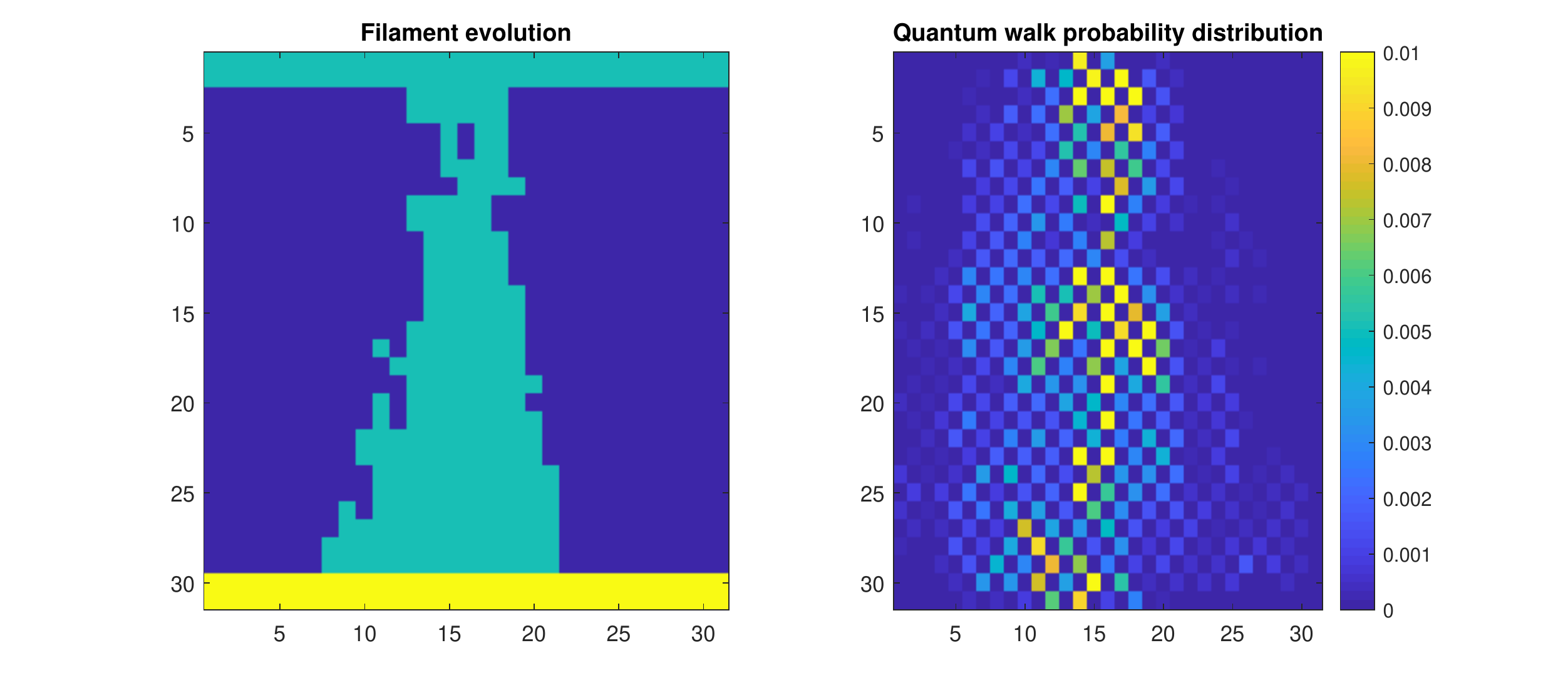}}\\
  \subfloat[]{\includegraphics[width=0.495\linewidth]{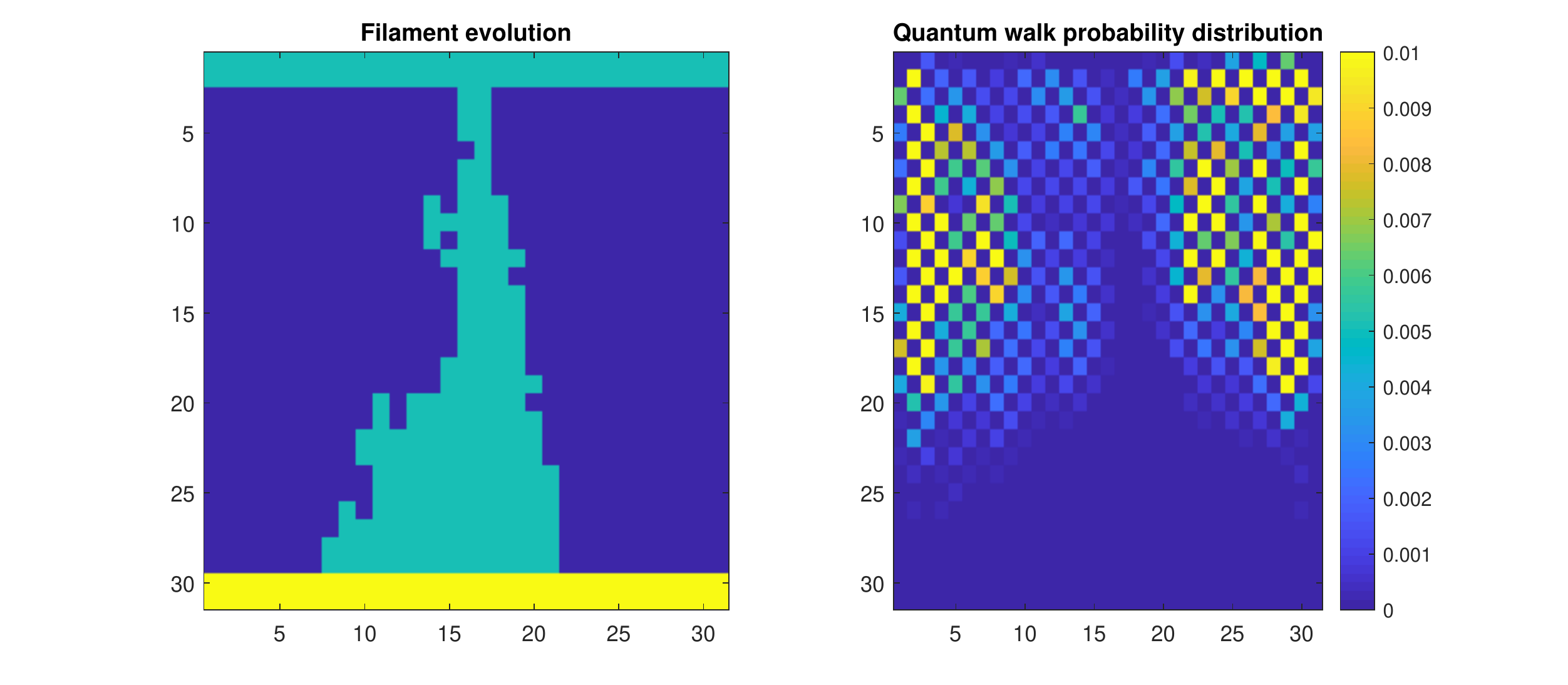}}
  \subfloat[]{\includegraphics[width=0.495\linewidth]{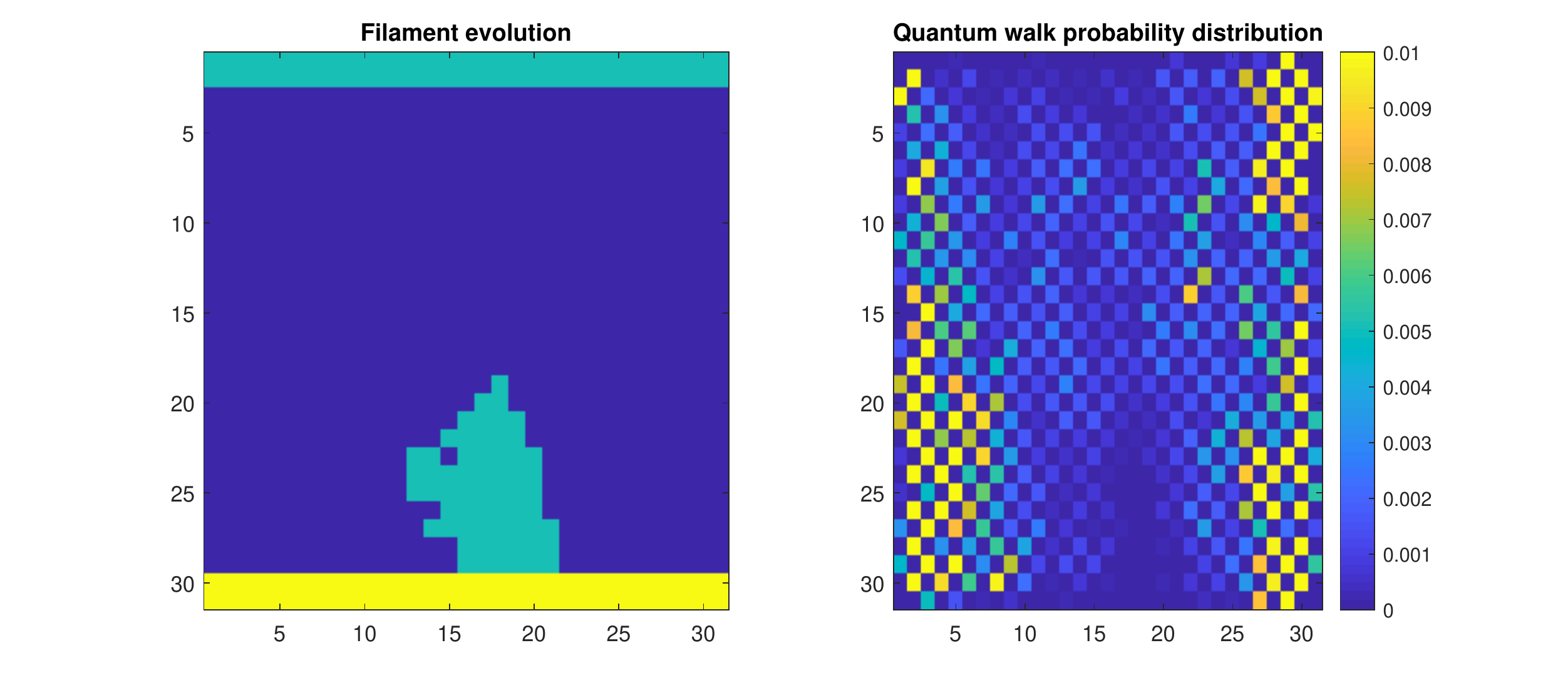}}
    \caption{Filament evolution due to Quantum Walker's probability distribution. (a) Initial filament growth with positive applied voltage. (b) Final filament formation. (c) Degradation of the filament due to the applied negative voltage. (d) Termination of the filament}
    \label{filament_evolution_qw}
\end{figure*}

In the above equation the symbol $i$, which is divided by the Plank constant $\hbar$ is the imaginary unit, while $V(i,j,s)$ is the potential at the $(i,j)$ cell at step $s$. The quantum walk models the kinetics of metallic ions in the following way: First the coin operator, $C(s)$, acts on the coin state, then the quantum walker (metallic ion) moves to one of the neighboring cells as a result of the shift operator, $P(s)$, acting on the previous location state and finally the product of these two operators is multiplied by the exponent of \eqref{eq17}, through which the potential distribution enters the computation. The coin operator $C(s)$ is the tensor product of two Hadamard quantum gates and by acting on the current coin state brings it to a superposition of the four basis coin states:

\begin{equation}
C(s)|\text{coin at time }{s}\rangle=c_R|00\rangle + c_L|01\rangle + c_U |10\rangle + c_D |11\rangle.
\label{eq18}
\end{equation}

In the above equation, $c_R$, $c_L$, $c_U$ and $c_D$ are the probability amplitudes, the absolute square of which gives the probability of the walker to move to the right, left, up and down, respectively. The shift operator acts on the location state and moves the quantum walker to one of the neighboring cells with probability determined by the probability amplitudes of eq.~(\ref{eq18}) as follows:

\begin{equation}
\label{eq19}
\begin{split}
P(t) = &\sum_{j=-n}^{i}|i+1,j\rangle,\langle i,j|\otimes{|00\rangle\langle{00}|}+\\
& |i-1,j\rangle,\langle i,j|\otimes{|01\rangle\langle{01}|}+|i,j+1\rangle,\langle i,j|\otimes{|10\rangle\langle{10}|}\\
& +|i,j-1\rangle,\langle i,j|\otimes{|11\rangle\langle{11}|}.
\end{split}
\end{equation}

To compute the next state of the quantum walk operator, \eqref{eq17} is applied to all cells of the 2D grid. The exponential factor containing the potential determines the constructive of destructive interference of the quantum walkers arriving at each cell, and the new distribution, at computational time step $s+1$, of the metallic ions into the dielectric is determined.

\begin{figure*}[!htbp]
    \centering
  \includegraphics[width=1\linewidth]{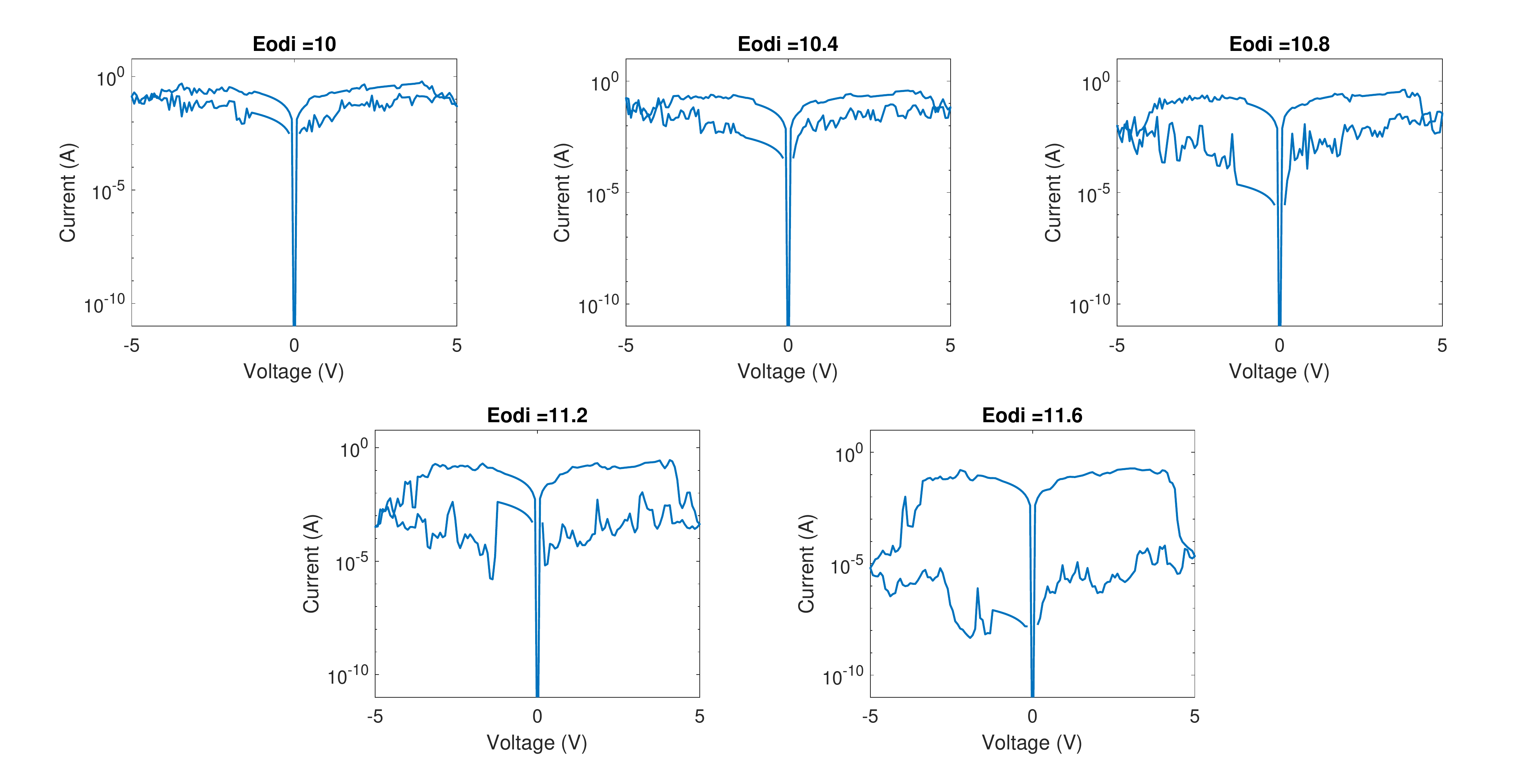}
    \caption{Logarithmic I-V for a range of dielectric's retention energy ($Eodi$) between 10eV and 11.6eV utilizing Quantum Walk model}
    \label{QW_IV_loga}
\end{figure*}

\begin{figure*}[!htbp]
    \centering
  \includegraphics[width=1\linewidth]{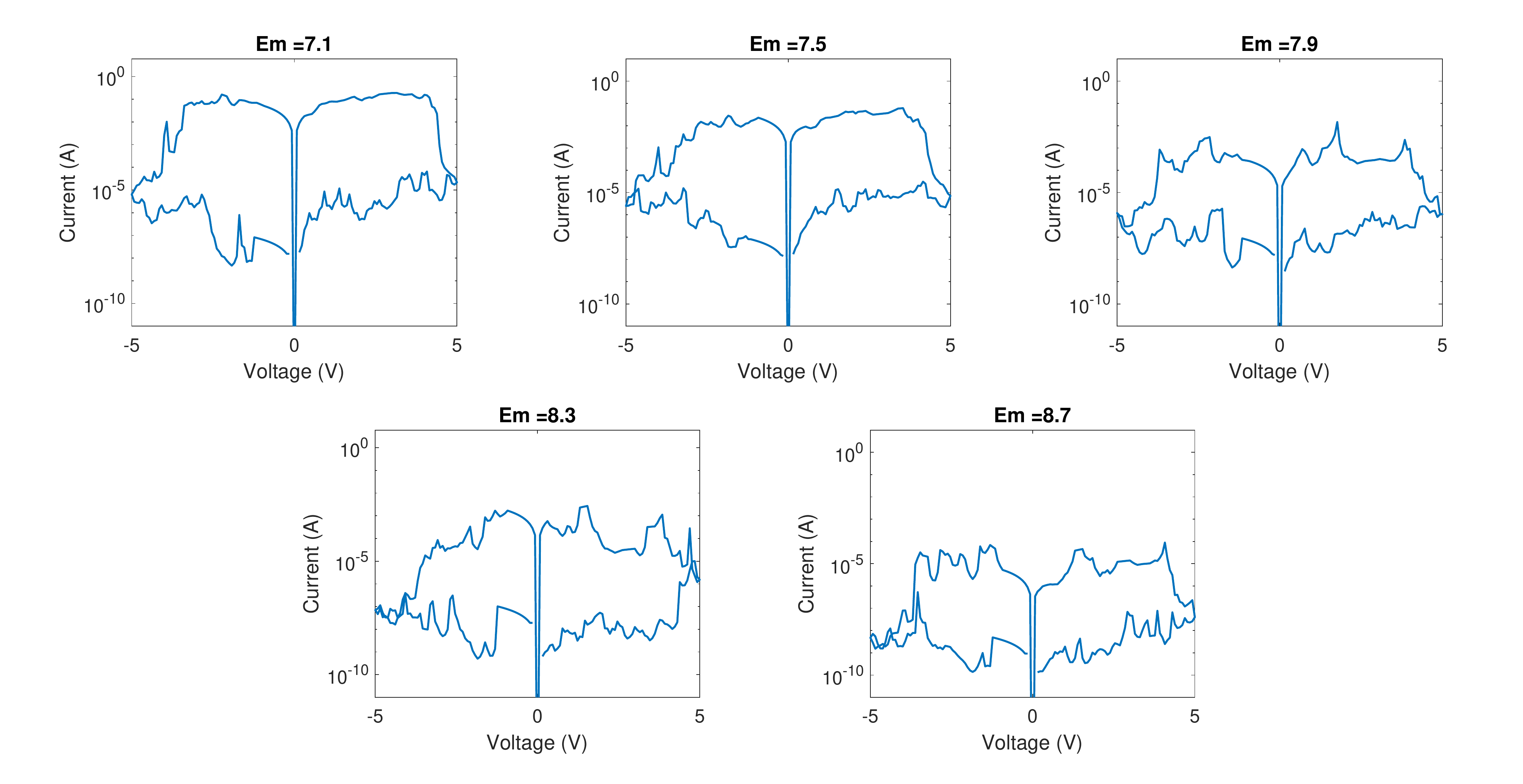}\\
    \caption{ Logarithmic I-V for a range of metal's retention energy ($Em$) between 7.1eV and 8.7eV utilizing Quantum Walk model}
    \label{QW_IV_logb}
\end{figure*}

\section{Simulation Results of Quantum-Walk based Filament formation}
\label{SimResults}

By executing the quantum walk algorithm the probabilities of the walker positions start to spread. In every computational time step of the evolution of the walker, a thresholding check is performed. The grid's locations whose probabilities surpass this threshold are occupied by a metallic ion. Consequently, a filament based on the evolution of the quantum walker is formed like the one in Fig.~\ref{filament_evolution_qw}(a,b). During negative applied voltage, two quantum walkers are initialized to the upper sides of the grid to simulate the degradation of the filament as shown in Figs.~\ref{filament_evolution_qw}(c,d), respectively. More specifically, due to the negative applied voltage, 
the probability of the backwards moving quantum walkers determines the probability of the metallic ions to move away from the previously formed filament. 
In Fig.~\ref{filament_evolution_qw}(c) it can be observed that while quantum walker's probability distribution surrounds the filament, metallic ions start to get removed and the filament becomes thinner.


In Fig.~\ref{QW_IV_loga} the logarithmic I-V curve for a different number of dielectric Retention Energies, namely $Eodi$, and in Fig.~\ref{QW_IV_logb} of metal Retention Energies, namely $Em$, are displayed, respectively. By examining Figs.~\ref{QW_IV_loga},~\ref{QW_IV_logb} it can be observed that when the dielectric retention energy is increased, the $R_{off}$ of the device becomes higher (vertically widening the I-V), while increasing the Metal retention energy both $R_{off}$ and $R_{on}$ increases (shifting the whole I-V down). These two variables can manipulate the electron motion and therefore, act as fitting parameters to regulate the $R_{off}$ and $R_{on}$ states of the fabricated devices.  

\begin{figure}[!t]
\centering
  \includegraphics[width=0.95\linewidth]{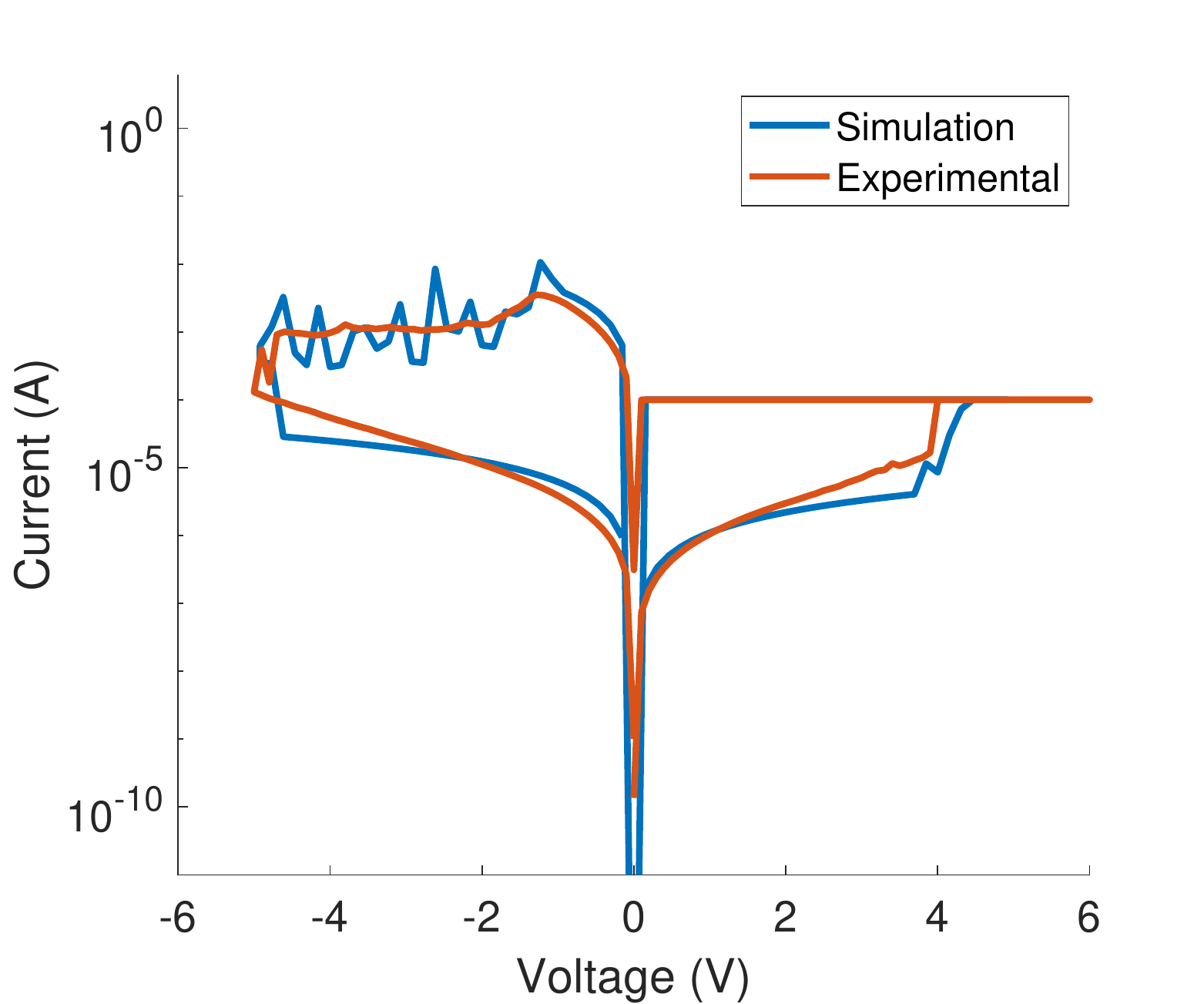}
    \caption{Comparison between experimental data and simulation output.}
    \label{ExperimentalvsModel}
\end{figure}

Finally, as a proof of concept of the proposed model abilities to simulate the operation of a filamentary resistive switching memristive device, a fitting of the presented model with a fabricated device has been conducted. As mentioned before, we fabricated and measured planar MIM devices made of 30nm Cu (active electrode) on a 40nm thick SiO$_2$ deposited by sputtering and 100nm thick Pt (counter electrode). Current - voltage staircase sweeps were performed in order to investigate the \texttt{SET} and \texttt{RESET} switching processes. The experimental results suggested a clear bipolar resistance switching operation with \texttt{SET} and \texttt{RESET} voltages~4.5V and -4.5V, respectively. The Cu/SiO$_2$ is a well-known combination to achieve ECM (MOxBF) ReRAM devices \cite{Tappertzhofen}. The resistance switching is attributed to the oxidation of Cu atoms lying at the interface with the oxide and the resulted Cu ions are drifted and diffuse across the bulk oxide layer towards the counter electrode.  More specifically, in Fig.~\ref{ExperimentalvsModel}, the comparison between the model and the experimental data can be thoroughly examined. The oscillation of the current during negative applied voltage is attributed to the scattering of the electrons due to disturbance of the material's lattice uniformity. This phenomenon is not present to the experimental data because the measuring setup takes multiple samples of the same voltage and, consequently, leading to a smoother I-V characteristic. Specifically, the I-V measurements were experimentally recorded using a semiconductor parameter analyzer HP4155A with medium integration time, that means the current was averaged over 1msec. 

A valuable further analysis that should be conducted is the time analysis of the filament's evolution. Evaluating the time interval of the filaments evolution parameter ($dt$) is required to specify the timing and frequency derived for the model, taking under consideration some physical variables in correspondence to the simulations carried out, such as the temperature distribution of the device, the electron mobility as well as the ion concentration. The ion mobility corresponds to the kinetic energy $t$ found in \eqref{eq1}, the ion concentration is linked to the number of atoms (size of the Hamiltonian) that constitute the device, as well as, to the quantum walker's distribution which is affected by the applied voltage and the internal temperature of the device is linked to the overlap integral.


\section{Conclusions}
\label{Conclusions}
In this work a novel quantum based model for filamentary resistive switching nanodevices, like MOxBF RS nanodevices, was presented. A quantum random walk based algorithm was developed to model the evolution of the conductive filament, while in each step of the filament’s evolution the conductance of the device is calculated by the GPU accelerated NEGF method. The simulation results provide clear identifications of the robustness of the model for a number of various retention energies for both dielectric and metal. Furthermore, as a single proof of concept, experimental data were used from a fabricated resistive switching device, namely planar MIM devices made of 30nm Cu (active electrode) on a 40nm thick SiO$_2$ deposited by sputtering and 100nm thick Pt (counter electrode) and were compared with the model's output when fitted to much the experimental data. The presented results showed a fair qualitative and quantitative agreement between the experimental and simulated data. As a further step, more parameters like temperature distribution, electron mobility and ion concentration of the under study devices will be considered. Finally, the GPU acceleration will be improved utilizing cuBLAS and cuSPARSE libraries to embed the efficiency of the sparse matrices into the GPU.

\section*{Acknowledgment}
We gratefully acknowledge the Greece-Russia bilateral joint research project MEM-Q (proj.no./MIS T4$\Delta$P$\Omega$-00030/5021467) supported by GSRT and funded by National and European funds.

\ifCLASSOPTIONcaptionsoff
  \newpage
\fi



\bibliographystyle{IEEEtran}
\bibliography{biblio}

\begin{thebibliography}{10}
\providecommand{\url}[1]{#1}
\csname url@samestyle\endcsname
\providecommand{\newblock}{\relax}
\providecommand{\bibinfo}[2]{#2}
\providecommand{\BIBentrySTDinterwordspacing}{\spaceskip=0pt\relax}
\providecommand{\BIBentryALTinterwordstretchfactor}{4}
\providecommand{\BIBentryALTinterwordspacing}{\spaceskip=\fontdimen2\font plus
\BIBentryALTinterwordstretchfactor\fontdimen3\font minus
  \fontdimen4\font\relax}
\providecommand{\BIBforeignlanguage}[2]{{%
\expandafter\ifx\csname l@#1\endcsname\relax
\typeout{** WARNING: IEEEtran.bst: No hyphenation pattern has been}%
\typeout{** loaded for the language `#1'. Using the pattern for}%
\typeout{** the default language instead.}%
\else
\language=\csname l@#1\endcsname
\fi
#2}}
\providecommand{\BIBdecl}{\relax}
\BIBdecl

\bibitem{chua2019handbook}
L.~Chua, G.~C. Sirakoulis, and A.~Adamatzky, \emph{Handbook of Memristor
  Networks}.\hskip 1em plus 0.5em minus 0.4em\relax Springer International
  Publishing, 2019.

\bibitem{dimitrakis21}
P.~Dimitrakis and E.~Valov, \emph{Metal Oxides for Non-volatile Memory}.\hskip
  1em plus 0.5em minus 0.4em\relax Elsevier, 2021.

\bibitem{edwards2015reconfigurable}
A.~H. Edwards, H.~J. Barnaby, K.~A. Campbell, M.~N. Kozicki, W.~Liu, and M.~J.
  Marinella, ``Reconfigurable memristive device technologies,''
  \emph{Proceedings of the IEEE}, vol. 103, no.~7, pp. 1004--1033, 2015.

\bibitem{Lanza}
M.~Lanza, H.-S.~P. Wong, E.~Pop, D.~Ielmini, D.~Strukov, B.~C. Regan,
  L.~Larcher, M.~A. Villena, J.~J. Yang, L.~Goux, A.~Belmonte, Y.~Yang, F.~M.
  Puglisi, J.~Kang, B.~Magyari-Köpe, E.~Yalon, A.~Kenyon, M.~Buckwell,
  A.~Mehonic, A.~Shluger, H.~Li, T.-H. Hou, B.~Hudec, D.~Akinwande, R.~Ge,
  S.~Ambrogio, J.~B. Roldan, E.~Miranda, J.~Suñe, K.~L. Pey, X.~Wu,
  N.~Raghavan, E.~Wu, W.~D. Lu, G.~Navarro, W.~Zhang, H.~Wu, R.~Li,
  A.~Holleitner, U.~Wurstbauer, M.~C. Lemme, M.~Liu, S.~Long, Q.~Liu, H.~Lv,
  A.~Padovani, P.~Pavan, I.~Valov, X.~Jing, T.~Han, K.~Zhu, S.~Chen, F.~Hui,
  and Y.~Shi, ``Recommended methods to study resistive switching devices,''
  \emph{Advanced Electronic Materials}, vol.~5, no.~1, p. 1800143, 2019.

\bibitem{Valov_2011}
I.~Valov, R.~Waser, J.~R. Jameson, and M.~N. Kozicki, ``Electrochemical
  metallization memories{\textemdash}fundamentals, applications, prospects,''
  \emph{Nanotechnology}, vol.~22, no.~25, p. 254003, may 2011.

\bibitem{Valov14}
I.~Valov, ``Redox-based resistive switching memories (rerams): Electrochemical
  systems at the atomic scale,'' \emph{ChemElectroChem}, vol.~1, 01 2014.

\bibitem{Menzel}
S.~Menzel, P.~Kaupmann, and R.~Waser, ``Understanding filamentary growth in
  electrochemical metallization memory cells using kinetic monte carlo
  simulations,'' \emph{Nanoscale}, vol.~7, pp. 12\,673--12\,681, 2015.

\bibitem{Menzel18}
S.~{Menzel}, A.~{Siemon}, A.~{Ascoli}, and R.~{Tetzlaff}, ``Requirements and
  challenges for modelling redox-based memristive devices,'' in \emph{2018 IEEE
  International Symposium on Circuits and Systems (ISCAS)}, 2018, pp. 1--5.

\bibitem{strukov2008missing}
D.~B. Strukov, G.~S. Snider, D.~R. Stewart, and R.~S. Williams, ``The missing
  memristor found,'' \emph{nature}, vol. 453, no. 7191, p.~80, 2008.

\bibitem{pickett2009switching}
M.~D. Pickett, D.~B. Strukov, J.~L. Borghetti, J.~J. Yang, G.~S. Snider, D.~R.
  Stewart, and R.~S. Williams, ``Switching dynamics in titanium dioxide
  memristive devices,'' \emph{Journal of Applied Physics}, vol. 106, no.~7, p.
  074508, 2009.

\bibitem{waser2009redox}
R.~Waser, R.~Dittmann, G.~Staikov, and K.~Szot, ``Redox-based resistive
  switching memories--nanoionic mechanisms, prospects, and challenges,''
  \emph{Advanced materials}, vol.~21, no. 25-26, pp. 2632--2663, 2009.

\bibitem{chang2011synaptic}
T.~Chang, S.-H. Jo, K.-H. Kim, P.~Sheridan, S.~Gaba, and W.~Lu, ``Synaptic
  behaviors and modeling of a metal oxide memristive device,'' \emph{Applied
  physics A}, vol. 102, no.~4, pp. 857--863, 2011.

\bibitem{jiang2016compact}
Z.~Jiang, Y.~Wu, S.~Yu, L.~Yang, K.~Song, Z.~Karim, and H.-S.~P. Wong, ``A
  compact model for metal--oxide resistive random access memory with experiment
  verification,'' \emph{IEEE Transactions on Electron Devices}, vol.~63, no.~5,
  pp. 1884--1892, 2016.

\bibitem{corinto2012boundary}
F.~Corinto and A.~Ascoli, ``A boundary condition-based approach to the modeling
  of memristor nanostructures,'' \emph{IEEE Transactions on Circuits and
  Systems I: Regular Papers}, vol.~59, no.~11, pp. 2713--2726, 2012.

\bibitem{joglekar2009elusive}
Y.~N. Joglekar and S.~J. Wolf, ``The elusive memristor: properties of basic
  electrical circuits,'' \emph{European Journal of Physics}, vol.~30, no.~4, p.
  661, 2009.

\bibitem{biolek2009spice}
Z.~Biolek, D.~Biolek, and V.~Biolkova, ``Spice model of memristor with
  nonlinear dopant drift.'' \emph{Radioengineering}, vol.~18, no.~2, 2009.

\bibitem{prodromakis2011versatile}
T.~Prodromakis, B.~P. Peh, C.~Papavassiliou, and C.~Toumazou, ``A versatile
  memristor model with nonlinear dopant kinetics,'' \emph{IEEE transactions on
  electron devices}, vol.~58, no.~9, pp. 3099--3105, 2011.

\bibitem{Ntinas}
V.~{Ntinas}, A.~{Ascoli}, R.~{Tetzlaff}, and G.~C. {Sirakoulis}, ``A complete
  analytical solution for the on and off dynamic equations of a {TaO}
  memristor,'' \emph{IEEE Transactions on Circuits and Systems II: Express
  Briefs}, vol.~66, no.~4, pp. 682--686, 2019.

\bibitem{kvatinsky2012team}
S.~Kvatinsky, E.~G. Friedman, A.~Kolodny, and U.~C. Weiser, ``Team: Threshold
  adaptive memristor model,'' \emph{IEEE Transactions on Circuits and Systems
  I: Regular Papers}, vol.~60, no.~1, pp. 211--221, 2012.

\bibitem{kvatinsky2015vteam}
S.~Kvatinsky, M.~Ramadan, E.~G. Friedman, and A.~Kolodny, ``{VTEAM}: A general
  model for voltage-controlled memristors,'' \emph{IEEE Transactions on
  Circuits and Systems II: Express Briefs}, vol.~62, no.~8, pp. 786--790, 2015.

\bibitem{vourkas2012novel}
I.~Vourkas and G.~C. Sirakoulis, ``A novel design and modeling paradigm for
  memristor-based crossbar circuits,'' \emph{IEEE Transactions on
  Nanotechnology}, vol.~11, no.~6, pp. 1151--1159, 2012.

\bibitem{datta2000nanoscale}
S.~Datta, ``Nanoscale device modeling: the green’s function method,''
  \emph{Superlattices and microstructures}, vol.~28, no.~4, pp. 253--278, 2000.

\bibitem{Tappertzhofen}
S.~Tappertzhofen, H.~Mündelein, I.~Valov, and R.~Waser, ``Nanoionic transport
  and electrochemical reactions in resistively switching silicon dioxide,''
  \emph{Nanoscale}, vol.~4, pp. 3040--3043, 2012.

\bibitem{Menzel15}
S.~Menzel, S.~Tappertzhofen, R.~Waser, and I.~Valov, ``Switching kinetics of
  electrochemical metallization memory cells,'' \emph{Phys. Chem. Chem. Phys.},
  vol.~15, pp. 6945--6952, 2013.

\bibitem{Valov_2017}
I.~Valov, ``Interfacial interactions and their impact on redox-based resistive
  switching memories ({ReRAMs}),'' \emph{Semiconductor Science and Technology},
  vol.~32, no.~9, p. 093006, Aug 2017.

\bibitem{Valov_2013}
I.~Valov and M.~N. Kozicki, ``Cation-based resistance change memory,''
  \emph{Journal of Physics D: Applied Physics}, vol.~46, no.~7, p. 074005, Feb
  2013.

\bibitem{Mehonic}
A.~Mehonic, A.~L. Shluger, D.~Gao, I.~Valov, E.~Miranda, D.~Ielmini,
  A.~Bricalli, E.~Ambrosi, C.~Li, J.~J. Yang, Q.~Xia, and A.~J. Kenyon,
  ``Silicon oxide (siox): A promising material for resistance switching?''
  \emph{Advanced Materials}, vol.~30, no.~43, p. 1801187, 2018.

\bibitem{Moysidis}
S.~Moysidis, I.~Karafyllidis, and P.~Dimitrakis, ``Design and simulation of
  graphene majority gate without back-gating,'' \emph{Engineering Research
  Express}, vol.~1, 08 2019.

\bibitem{Moysidis18}
S.~{Moysidis}, I.~G. {Karafyllidis}, and P.~{Dimitrakis}, ``Graphene logic
  gates,'' \emph{IEEE Transactions on Nanotechnology}, vol.~17, no.~4, pp.
  852--859, 2018.

\bibitem{aharonov1993quantum}
Y.~Aharonov, L.~Davidovich, and N.~Zagury, ``Quantum random walks,''
  \emph{Physical Review A}, vol.~48, no.~2, p. 1687, 1993.

\end{thebibliography}
%


\end{document}